\documentclass[sigconf, nonacm]{acmart}
\usepackage{trec-style}
\AtBeginDocument{%
  \providecommand\BibTeX{{%
    \normalfont B\kern-0.5em{\scshape i\kern-0.25em b}\kern-0.8em\TeX}}}

\begin{document}
\settopmatter{printacmref=false}
\title{Generative Relevance Feedback and Convergence of Adaptive Re-Ranking: University of Glasgow Terrier Team at TREC DL 2023}

\author{Andrew Parry}
\email{a.parry.1@research.gla.ac.uk}
\orcid{1234-5678-9012}
\affiliation{%
  \institution{University of Glasgow}
  \city{Glasgow}
  \country{UK}
}
\author{Thomas J\"{a}nich}
\email{t.jaenich.1@research.gla.ac.uk}
\affiliation{%
  \institution{University of Glasgow}
  \city{Glasgow}
  \country{UK}
}

\author{Sean MacAvaney}
\author{Iadh Ounis}
\email{first.last@glasgow.ac.uk}
\affiliation{%
  \institution{University of Glasgow}
  \city{Glasgow}
  \country{UK}
}
\renewcommand{\shortauthors}{Parry, et al.}

\begin{abstract}
    This paper describes our participation in the TREC 2023 Deep Learning Track. We submitted runs that apply generative relevance feedback from a large language model in both a zero-shot and pseudo-relevance feedback setting over two sparse retrieval approaches, namely BM25 and SPLADE. We couple this first stage with adaptive re-ranking over a BM25 corpus graph scored using a monoELECTRA cross-encoder. We investigate the efficacy of these generative approaches for different query types in first-stage retrieval. In re-ranking, we investigate operating points of adaptive re-ranking with different first stages to find the point in graph traversal where the first stage no longer has an effect on the performance of the overall retrieval pipeline. We find some performance gains from the application of generative query reformulation. However, our strongest run in terms of P@10 and nDCG@10 applied both adaptive re-ranking and generative pseudo-relevance feedback, namely \textsf{uogtr\_b\_grf\_e\_gb}.
\end{abstract}

\maketitle

\section{Introduction}

The University of Glasgow Terrier team participated in the TREC
2023 Deep Learning track to further explore new generative approaches to retrieval and validate existing approaches by a deeper exploration of their performance. We investigate generative approaches to relevance feedback utilising both generative query reformulation (Gen-QR) and pseudo-relevance feedback (Gen-PRF)~\cite{genqr} as well as conducting a further evaluation of adaptive re-ranking~\cite{gar} on the MS MARCO-v2 corpus~\cite{msmarco}. Both of these approaches overcome the lexical mismatch problem of many classic retrieval approaches, albeit in different ways. A re-ranker is normally constrained by the texts that could be retrieved by a first-stage model. Generative relevance feedback alleviates this problem by adding expansion terms to the query text attempting to improve the recall of first-stage retrieval models. Adaptive re-ranking instead expands the initial ranking with candidate documents found by traversal of a corpus graph to find the nearest neighbours of highly ranked texts.

We explored the following research questions: (1) Can we validate the performance improvements from generative relevance feedback on a different test set and downstream ranking model? (2) What type of queries are most improved or harmed by generative relevance feedback? and (3) Can the use of a sufficiently large corpus graph allow a lightweight first-stage ranker to converge to identical performance of a first-stage learned retrieval function?

To answer these questions, we used our PyTerrier Information Retrieval (IR) toolkit~\cite{terrier}, which allows for the composition of both lexical and neural retrieval components into flexible pipelines. We apply two forms of generative relevance feedback over two sparse first-stage ranking models. Moreover, we apply adaptive re-ranking with monoELECTRA~\cite{monoelectra} in this setting to assess the need for complex first-stage pipelines (learned models or weighted expansion terms) given a sufficiently large corpus graph. Specifically, we use a large search budget of 5000 over a 32 nearest-neighbour BM25 corpus graph.

The structure of the remainder of this paper is as follows: Section \ref{sec:pipe} details the notation of retrieval pipelines composed in PyTerrier. Section \ref{sec:method} describes the methods used in our experiments and the particular models used in these methods. We then outline our experiment setup. Section \ref{sec:runs} summarises our submitted baseline and group runs. In Section \ref{sec:results} we present our results and analysis before providing some concluding remarks in Section \ref{sec:conclus}.

\section{Pyterrier Retrieval Pipelines}
\label{sec:pipe}
Our experiments and submitted runs for the TREC 2023 Deep Learning Track have been built upon  PyTerrier, Python bindings for the Terrier search engine~\cite{terrier}. The key abstraction of PyTerrier is the \textit{Transformer} object, of which all retrieval components are a sub-class. Using Pandas dataframes, a transformer object \textit{transforms} one dataframe to another. To create multi-stage retrieval pipelines, PyTerrier overloads the bitwise shift operators (<< and >>) to allow the chaining of multiple transformers into a single component where the output of each transformer is directly passed into the next sequentially.

We express all ranking features described in the rest of this paper as pipelines of transformers using this sequential operator. For more information about the PyTerrier platform and the operators' flexibility, we refer the reader to the documentation, which can be found at \url{https://pyterrier.readthedocs.io/en/latest/}.

\section{Methods}
\label{sec:method}
In this section, we outline the background knowledge of Generative Relevance Feedback (Section \ref{sec:feedback}) and Adaptive Re-Ranking (Section \ref{sec:gar}), followed by our experimental setup in Section \ref{sec:setup}.

\subsection{Generative Query Reformulation \& Pseudo-Relevance Feedback}
\label{sec:feedback}
Generative query reformulation (Gen-QR) and generative pseudo-relevance feedback (Gen-PRF) are mechanisms for the expansion of query terms in ranking pipelines~\cite{genqr}. Originally proposed with both a weakly supervised fine-tuned variant as well as a zero-shot prompted instruction-tuned variant, we only cover the latter as we do not utilise any fine-tuned checkpoints in our generative expansion stage.

Recent developments in language model pre-training~\cite{radford, llama, taori_stanford_2023} have led to a new paradigm in language modelling known as prompting~\cite{prompt} in which a task is described with some input in a zero-shot setting to a language model. Due to extensive pre-training and additional fine-tuning on instruction style inputs~\cite{flan}, these models can perform a task that has not been observed in training data~\cite{prompt}. This zero-shot generalisation poses some benefits in retrieval as the labelling of ground truth by experts for neural models is an expensive process~\cite{expense}.

Following \citet{genqr}, we define query reformulation as a process $\mathcal{P}$ which, given a user query $q^0$, reformulates it as $q^r$ potentially conditioned on some additional context to improve the retrieval effectiveness of the downstream retrieval pipeline.

\begin{equation}
    q^r = \mathcal{P}(q^0, \dots)
    \label{eq:reform}
\end{equation}

In equation \ref{eq:reform}, $\dots$ is any additional context, such as an initial ranking in the case of pseudo-relevance feedback. Generative relevance feedback prompts a large language model (LLM) to suggest expansion terms to improve the performance of a retrieval model~\footnote{Prompt for \textsf{Gen-PRF}: Improve the search effectiveness by suggesting expansion terms for the query:\{input\_query\}, based on the given context information: \{context\}}. In the case of Gen-QR, no additional context is supplied to the generative model. However, when applying Gen-PRF, a ranking of $K$ documents from a corpus $D$ by some score function $s$, \{$s(d, q^0); \forall d \in D$\} is provided as topical context for the expansion and/or re-weighting of query terms.

The authors propose multiple pruning criteria to reduce the context size provided to the Gen-PRF model. We use the approach \textit{Top-P} and outline its method as follows. To select $M$ context pieces from documents in an initial ranking $R$ (this process is simplified for passage ranking as we do not further split passages), each document $d$ is split into passages. The context $C$ of length $M$ is then chosen as follows:

\begin{equation}
    C_\textit{Top-P}(R, q^0) = \argmax_{p \in R}^M \ s \ s(p, q^0)
\end{equation}

Example pipelines are as follows: Gen-QR with FLAN-T5 can be applied before BM25 retrieval, re-ranked with monoELECTRA:

\begin{equation}
    \textsf{QR}(\textsf{FLAN-T5}) >> \textsf{BM25} >> \textsf{monoELECTRA}
    \label{eq:qr}
\end{equation}

Where in Equation \ref{eq:qr}, $\textsf{QR}(\textsf{FLAN-T5})$ is the generative reformulation of a query using FLAN-T5. Alternatively, Top-P context Gen-PRF with FLAN-T5 can be applied to a first-stage retrieval by BM25, re-ranked by monoELECTRA:

\begin{equation}
    \textsf{BM25} >> \textsf{PRF}_\textit{Top-P}(\textsf{FLAN-T5}) >> \textsf{BM25} >> \textsf{monoELECTRA}
\end{equation}

\subsection{Graph-based Adaptive Re-Ranking (GAR)}
\label{sec:gar}
We utilise graph-based adaptive re-ranking (GAR), which is an efficient approach to improving re-ranking performance~\cite{gar}. GAR works within a cascading retrieval pipeline in which new candidate documents are found by traversing a nearest-neighbour graph $G=(V, E)$ where each node in $V$ represents a document in the corpus. Each edge in $E$ is weighted by some heuristic, either a lexical or semantic similarity score between two nodes. Online latency is unaffected by the chosen number of nearest neighbours. However, due to the graph structure's quadratic time and space complexity, we limit the number of nearest neighbours to a small number and prune by sorting edge weights in descending order. An online traversal of this corpus graph is performed by scoring the nearest neighbours of the initial ranking using some score model $S$. The highest-ranking document neighbours from the current corpus graph frontier are added to a re-ranking pool. The initial pool is then revisited to update the graph frontier. This process alternates between scoring pools until a predefined compute budget is met.

An example pipeline is as follows: A first-stage BM25 retrieval is adaptively re-ranked using GAR with a BM25 corpus graph and a monoELECTRA scoring model: 

\begin{equation}
    \textsf{BM25} >> \textsf{GAR}_{\textsf{BM25}}(\textsf{monoELECTRA})
    \label{eq:gar}
\end{equation}

Where in Equation \ref{eq:gar}, $\textsf{GAR}_{\textsf{BM25}}(\textsf{monoELECTRA})$ represents adaptive re-ranking over a BM25 corpus graph using monoELECTRA.

\subsection{Experimental Setup}
\label{sec:setup}
We use the following retrieval components, grouped into methods that perform retrieval or re-ranking and methods that perform query expansion in the form of query reformulation (QR), pseudo-relevance feedback (PRF), or adaptive re-ranking (GAR). 

\para{Retrieval:}
\uls 
\li \textsf{DPH~\cite{dph} \& BM25~\cite{bm25}}:  Lexical retrieval from a Terrier inverted
index over the msmarco-passage-v2 corpus.
\li \textsf{SPLADE}\footnote{\textsf{naver/splade-cocondenser-ensembledistil}}: A distilled SPLADE++ learned sparse model~\cite{splade, splade2}.
\li \textsf{monoELECTRA}\footnote{\textsf{crystina-z/monoELECTRA\_LCE\_nneg31}}: A cross-encoder pre-trained with ELECTRA-style objectives~\cite{electra} and fine-tuned with 31 localized negatives~\cite{monoelectra}.
\ule 

\para{Query Expansion:}
\uls 
\li \textsf{Bo1} : Pseudo-relevance feedback using the DFR Bo1 model~\cite{amati_probability_2003}
over a Terrier index.
\li \textsf{Gen-QR (QR(LLM))}: Using the generative encoder-decoder FLAN-T5\footnote{\textsf{google/flant5-xxl}}~\cite{flan} to provide zero-shot query expansion terms with term weights controlled uniformly by the parameter $\beta=0.5$. In all cases, we use tuned generation hyperparameters from \citet{genqr}. 
\li \textsf{Gen-PRF (PRF$_C$(LLM))}: Using the generative encoder-decoder FLAN-T5 to provide query expansion terms using a first-stage ranking as context $C$. Term weights are again controlled uniformly by the parameter $\beta=0.5$. 
\li \textsf{GAR}$_G(S)$ : Graph-based Adaptive Re-Ranking using corpus
graph $G$ and scoring function $S$~\cite{gar}. We use a BM25-based corpus graph with a re-ranking budget of 5000 and 32 nearest neighbours.
\ule 

All experiments were conducted in PyTerrier using the ir-datasets package for corpora and additional test sets (DL-2021~\cite{dl21}, DL-2022~\cite{dl22}). PyTerrier is available from \url{https://github.com/terrier-org/pyterrier} and the implementation of monoELECTRA is available from \url{https://github.com/terrierteam/pyterrier_dr}. The implementation of GAR is available from \url{https://github.com/terrierteam/pyterrier_adaptive} and the implementation of Generative QR and PRF are available from \url{https://github.com/Parry-Parry/pyterrier_GenerativeQR}. We perform all experiments on a de-duped msmarco-v2 corpus due to the added benefits for GAR corpus graph traversal preventing duplicate nearest neighbours. We then added duplicates before submission. All experiment code is available from \url{https://github.com/Parry-Parry/terrier_trec_2023}.

\section{Submitted Runs}
\label{sec:runs}
We submitted six group runs to the passage ranking task. We also
submitted five baseline runs. We did not participate in the document
ranking task.

\subsection{Baseline Runs}

As baseline runs submitted to the 2023 Deep Learning passage ranking track, we chose two sparse retrieval models, one learned and one non-parametric model, one sparse retrieval model with query expansion and two sparse retrieval models re-ranked by a monoELECTRA cross-encoder. All baselines are summarised as follows:

\uls 
\li \textsf{uofg\_tr\_dph}:  Performs DPH retrieval on our passage sparse index.
\li \textsf{uofg\_tr\_s} : Performs SPLADE retrieval retrieval on an msmarco-passage-v2
SPLADE learned sparse index.
\li \textsf{uofg\_tr\_dph\_bo1}: Performs Bo1 divergence from randomness query expansion over DPH retrieval on our passage sparse index.
\li \textsf{uofg\_tr\_be}: Re-Ranks a first-stage BM25 retrieval using a monoELECTRA cross-encoder.
\li \textsf{uofg\_tr\_se}: Re-Ranks a first-stage SPLADE retrieval using monoELECTRA.
\ule

\subsection{Submitted Group Runs}

For the 2023 Deep Learning passage ranking track, we submitted
the following three group runs:

\uls 
\li \textsf{uofg\_tr\_se\_gb}: Performs SPLADE retrieval, adaptively re-ranked
by monoELECTRA and a BM25 graph.
\li \textsf{uofg\_tr\_qr\_be\_gb}: Performs Generative Query Reformulation before BM25 retrieval, adaptively re-ranked by monoELECTRA and a BM25 graph.
\li \textsf{uofg\_tr\_b\_grf\_e\_gb}: Performs Generative Relevance Feedback over a first-stage BM25 retrieval re-ranked by monoELECTRA and a BM25 graph.
\ule

\subsection{Additional Runs}

For the 2023 Deep Learning passage ranking track, we submitted
the following additional three runs:

\uls 
\li \textsf{uofg\_tr\_qr\_be}: Performs Generative Query Reformulation before BM25 retrieval re-ranked by monoELECTRA.
\li \textsf{uofg\_tr\_b\_grf\_e}: Performs Generative Relevance Feedback over a first-stage BM25 retrieval re-ranked by monoELECTRA.
\li \textsf{uofg\_tr\_be\_gb}: Performs BM25 retrieval, adaptively re-ranked
by monoELECTRA and a BM25 graph.
\ule

\section{Results \& Analysis}
\label{sec:results}
\begin{table*}[t]
\centering
\setlength{\tabcolsep}{2pt}
\caption{Results on the TREC Deep Learning 2023 Passage Ranking task. Best and Median performance is an aggregate over the best-performing model for each topic. Runs that outperform the median are denoted with $\better$. The strongest performance in each metric is denoted \textbf{bold}, the second strongest is \underline{underlined}.}
\label{tab:main}
\begin{tabular}{@{}ll|ccccc@{}}
\toprule
\textbf{Run ID} & \textbf{Pipeline} & \textbf{P@10} & \textbf{NDCG@10} & \textbf{MRR} & \textbf{MAP} & \textbf{R@100} \\
\midrule
\multicolumn{2}{l|}{Best (Per-topic aggregate)} & $0.7000\worse$ & $0.7892\worse$ & $0.9939\worse$ & $0.3839\worse$ & - \\
\multicolumn{2}{l|}{Median (Per-topic aggregate)} & $0.4085\worse$ & $0.5329\worse$ & $0.7803\worse$ & $0.2159\worse$ & - \\
\midrule
\multicolumn{7}{l}{\textbf{Baseline Runs}}\\
\midrule
\textsf{uogtr\_dph} & \textsf{DPH} &$0.1805\worse$ & $0.2825\worse$ & $0.4320\worse$ & $0.0840\worse$ & $0.2310\worse$ \\
\textsf{uogtr\_s} & \textsf{SPLADE} &$0.3549\worse$ & $0.4706\worse$ & $0.6725\worse$ & $0.1925\worse$ & $0.4144\worse$ \\
\textsf{uogtr\_dph\_bo1} & \textsf{DPH >> Bo1 >> DPH} &$0.1317\worse$ & $0.2377\worse$ & $0.3613\worse$ & $0.0600\worse$ & $0.1288\worse$ \\
\textsf{uogtr\_be} & \textsf{BM25 >> monoELECTRA} &$0.3939\worse$ & $0.5227\worse$ & $0.7881\better$ & $0.1940\worse$ & $0.3558\worse$ \\
\textsf{uogtr\_se} & \textsf{SPLADE >> monoELECTRA} &$0.4256\better$ & $0.5364\better$ & $0.7938\better$ & $\underline{0.2348}\better$ & $\underline{0.4531}\worse$ \\
\midrule
\multicolumn{7}{l}{\textbf{Additional Runs}} \\
\midrule
    \textsf{uogtr\_qr\_be} & \textsf{QR(FLAN-T5) >> BM25 >> monoELECTRA} &$0.3963\worse$ & $0.5316\worse$ & $\underline{0.8018}\better$ & $0.1994\worse$ & $0.3668\worse$ \\
\textsf{uogtr\_b\_grf\_e} & \textsf{BM25 >> PRF$_{\textit{Top-P}}$(FLAN-T5) >> BM25 >> monoELECTRA }&$0.4122\better$ & $0.5376\better$ & $0.7790\worse$ & $0.1996\worse$ & $0.3770\worse$ \\
\textsf{uogtr\_be\_gb} & \textsf{BM25 >> GAR$_{BM25}$(monoELECTRA)} &$0.4159\better$ & $0.5451\better$ & $\textbf{0.8046}\better$ & $0.2285\better$ & $0.4240\worse$ \\
\midrule
\multicolumn{7}{l}{\textbf{Group Runs}}\\
\midrule
\textsf{uogtr\_se\_gb} & \textsf{SPLADE >> GAR$_{BM25}$(monoELECTRA)} &$\underline{0.4293}\better$ & $0.5394\better$ & $0.7934\better$ & $\textbf{0.2401}\better$ & $\textbf{0.4706}\worse$ \\
\textsf{uogtr\_qr\_be\_gb} & \textsf{QR(FLAN-T5) >> BM25 >> GAR$_{BM25}$(monoELECTRA)} &$0.4244\better$ & $\underline{0.5488}\better$ & $0.7953\better$ & $0.2315\better$ & $0.4248\worse$ \\
\textsf{uogtr\_b\_grf\_e\_gb} & \textsf{BM25 >> PRF$_{\textit{Top-P}}$(FLAN-T5) >> BM25 >> GAR$_{BM25}$(monoELECTRA)}  &$\textbf{0.4305}\better$ & $\textbf{0.5489}\better$ & $0.7885\better$ & $0.2314\better$ & $0.4328\worse$ \\
\bottomrule
\end{tabular}
\end{table*}

In Table \ref{tab:main}, we present the performance of each of our runs contrasting group runs with baseline runs\footnote{Due to an issue with indexing, expansion terms degrade DPH (\textsf{uogtr\_dph\_bo1}).} and the Best and Median results averaged across all judged topics. Across all metrics, our group runs improve over the median, with our additional runs also improving over the median in some cases, notably when adaptive re-ranking is applied with monoELECTRA over BM25 (\textsf{uogtr\_be\_gb}). 

\para{Generative methods versus SPLADE}
Our SPLADE baseline re-ranked with monoELECTRA (\textsf{uogtr\_se}) outperforms all runs with BM25 first stages in terms of Recall@100 and MAP even when adaptive re-ranking is applied and outperforms all Gen-QR runs in terms of nDCG@10. As generative relevance feedback is a completely zero-shot method, this is not unexpected and consistent performance improvements are observed across all metrics. Comparing \textsf{uogtr\_be} and \textsf{uogtr\_qr\_be}, the diversification of query terms by Gen-QR is effective in improving first-stage lexical ranking models but fails to outperform the expansions of the learned SPLADE model. We observe improvements over SPLADE when applying Gen-PRF in nDCG@10, MAP, and R@100, suggesting that the context provided from the first-stage ranking is sufficient to ground expansion terms to suitable topics.

\para{GAR is a stronger standalone method}
Contrasting the use of generative relevance feedback (\textsf{uogtr\_qr\_be} and \textsf{uogtr\_b\_grf\_e}) against the use of GAR (\textsf{uogtr\_be\_gb}) we observe that GAR alone provides stronger candidates to the re-ranking stage than Gen-QR or Gen-PRF improving both precision and recall based metrics. Furthermore, measured by 4 of 5 metrics, the use of GAR is more effective in improving performance than the use of SPLADE expansions. We propose that this is due to the rank bias of the GAR algorithm, as candidate texts are chosen from the neighbourhood of the highest-ranking documents aligning well with the cluster hypothesis in contrast to the zero-shot expansion terms provided by generative relevance feedback. However, the computational expense of GAR is determined by a compute budget, and further research is warranted to find the Pareto-optimal approach for smaller budgets.

\para{Gen-PRF with GAR is most effective}
Our most effective run by P@10 and nDCG@10 uses Gen-PRF and adaptive re-ranking. We observe improvements in nDCG@10 over SPLADE expansions (0.5489 versus 0.5394) when using GAR, suggesting that the diversification provided by generative expansion terms can be more effective in providing a strong initial pool for GAR corpus graph traversal. We note that SPLADE-based runs (\textsf{uogtr\_se} and \textsf{uogtr\_se\_gb}) have greater recall and MAP than any generative relevance feedback pipelines. However, the zero-shot nature of these approaches coupled with user tendencies to only interact with the top 10 results of a search~\cite{perpage} reinforces generative relevance feedback as a compelling approach compared to learned first-stage models. Furthermore, addressing RQ (1), we find that Gen-QR and Gen-PRF generalise to a new test set and improve the retrieval performance of a monoELECTRA-based pipeline with and without adaptive re-ranking.

\subsection{Generative Relevance Feedback}

\begin{figure*}[t]
    \vspace*{-.7cm}
    \centering
    \subfloat[][Open Directed Queries]{\includegraphics[width=0.33\textwidth]{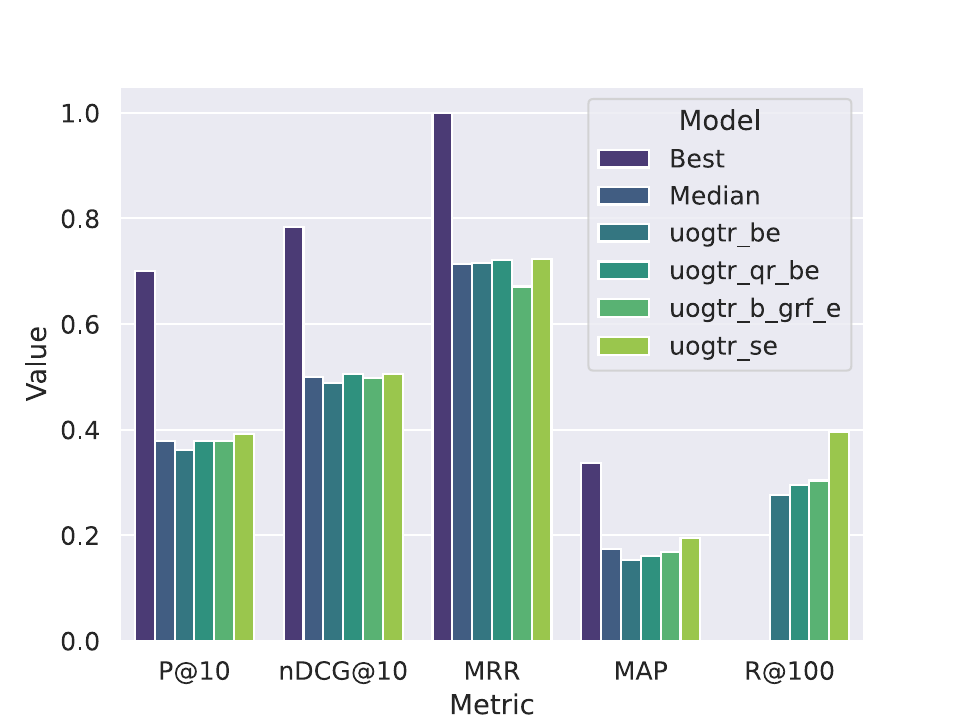}
    }
   \subfloat[][Closed Directed Queries]{\includegraphics[width=0.33\textwidth]{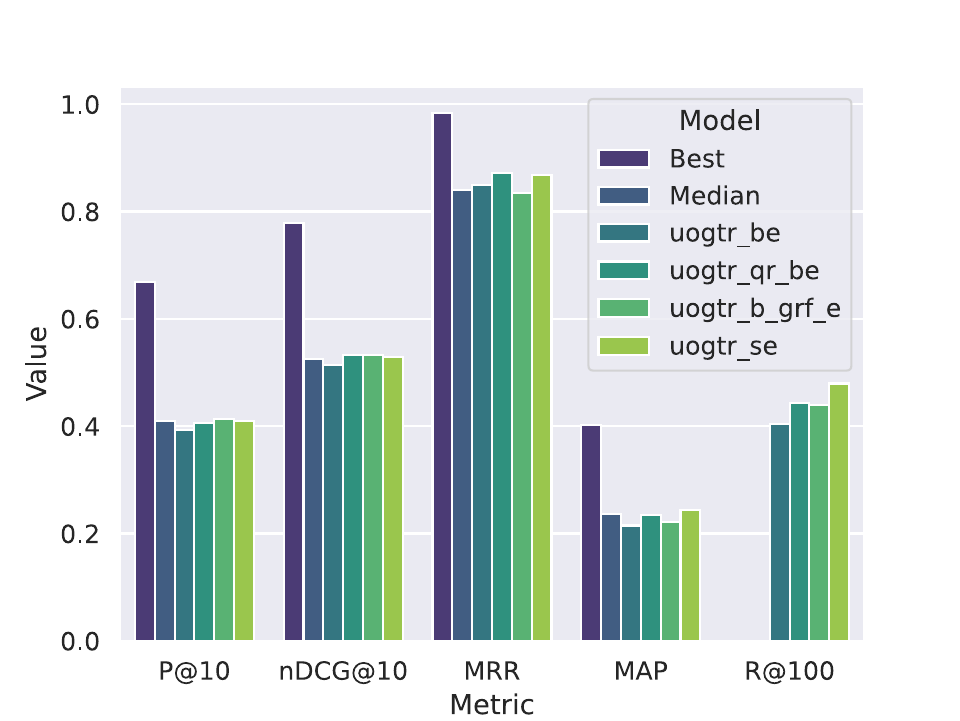}
    }
    \subfloat[][Advice Queries]{\includegraphics[width=0.33\textwidth]{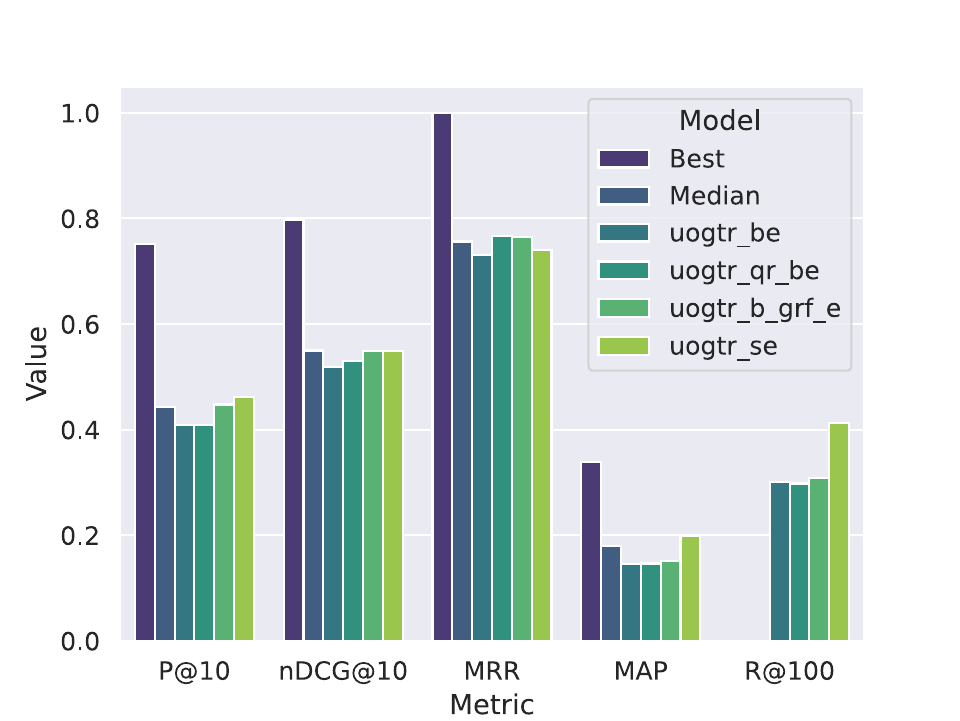}
    }
    \caption{Results for each main query type comparing \textsf{BM25 >> monoELECTRA}, \textsf{QR(FLAN-T5) >> BM25 >> monoELECTRA}, \textsf{BM25 >> PRF$_\textit{Top-P}$(FLAN-T5) monoELECTRA} and \textsf{SPLADE >> monoELECTRA}.}
    \label{fig:types}
\end{figure*}

To assess where generative relevance feedback is most effective, we compare generative expansions with learned expansions via SPLADE as well as a standard re-ranking pipeline across different query types. We refer users to the query type definitions proposed by Broder~\cite{taxonomy} and expanded by \citet{usergoals}. We focus on directed queries and advice, as the majority of test queries can be grouped into these classes. As defined by \citet{usergoals}, directed queries aim to learn a particular piece of information about a topic. They can be either closed when there is "a single, unambiguous answer" to a query or open when the query is an "open-ended question or one with unconstrained depth". Advice queries are requests for "advice, ideas, suggestion or instructions". We manually classified each Deep Learning 2023 test query into the groupings mentioned above. From 82 queries, we found that 31 fell in the directed closed class, 23 in the directed open class and 21 in the advice class. We investigate the performance of our runs within these query groupings.

As shown in Figure \ref{fig:types} (a), in open-directed queries, we find that Gen-QR is more effective than Gen-PRF or SPLADE in terms of nDCG@10. As the zero-shot expansions of Gen-QR are not grounded by either training in-domain or few-shot context from a first-stage ranker we hypothesize that in cases where the information need is broad, the relatively unconstrained diversification provided by such expansion terms leads to an improvement in both P@10 and nDCG@10 over other methods. A SPLADE first-stage retrieval (\textsf{uogtr\_se}) shows the strongest performance in terms of MRR, MAP, and R@100, suggesting that the in-domain training of SPLADE allows for the retrieval of more partially relevant documents than a generative approach.

In Figure \ref{fig:types} (b) we present performance on closed-directed queries. We observe that any linear trend in metric performance across all official evaluation metrics observed in Table \ref{tab:main} does not hold for queries with an unambiguous intent. We consider that diversification of terms may be unnecessary in such cases with Gen-PRF (\textsf{uogtr\_b\_grf\_e}) actually degrading performance over a standard monoELECTRA re-ranker (\textsf{uogtr\_be}) in terms of MRR. Furthermore, we observe that in terms of R@100 the grounding of Gen-PRF using first-stage ranking context does not improve recall over Gen-QR, reinforcing that pseudo-relevance feedback is unnecessary in these cases.

In Figure \ref{fig:types} (c) we present performance on advice queries. We observe that Gen-QR generally performs worse than Gen-PRF and SPLADE in terms of P@10 and nDCG@10. Upon inspection of generated expansions by Gen-QR, we observe that without either fine-tuning or first-stage context to ground expansion terms, the LLM directly generated advice instead of expansion terms. As we removed stopwords post-generation, much of the semantic structure was not used. However, many terms formed a part of an answer as opposed to more general terms for expansion. Inherently this is an artefact of the instruction-tuning of FLAN-T5, it does not appear to affect search more generally, as we still observe strong performance across other query types. However, this finding may explain the margin between Gen-QR and Gen-PRF. In future applications, either fine-tuning or the use of a stronger LLM may alleviate this issue. In the case of question-style or conversational-style queries, the LLM misunderstands the full instruction and instead attempts to answer the question directly which is not always conducive to the generation of effective expansion terms.

Concerning RQ (2), we find that generative relevance feedback can improve directed queries with both clear and ambiguous intent. However, performance can be harmed when a query is posed as a conversational style question, as the underlying LLM can attempt to answer the question instead of generating suitable expansion terms.

\subsection{Convergence of Adaptive Re-Ranking}

\begin{figure*}[t]
    \vspace*{-.7cm}
    \centering
    \subfloat[][RBO Correlation]{\includegraphics[width=0.5\textwidth]{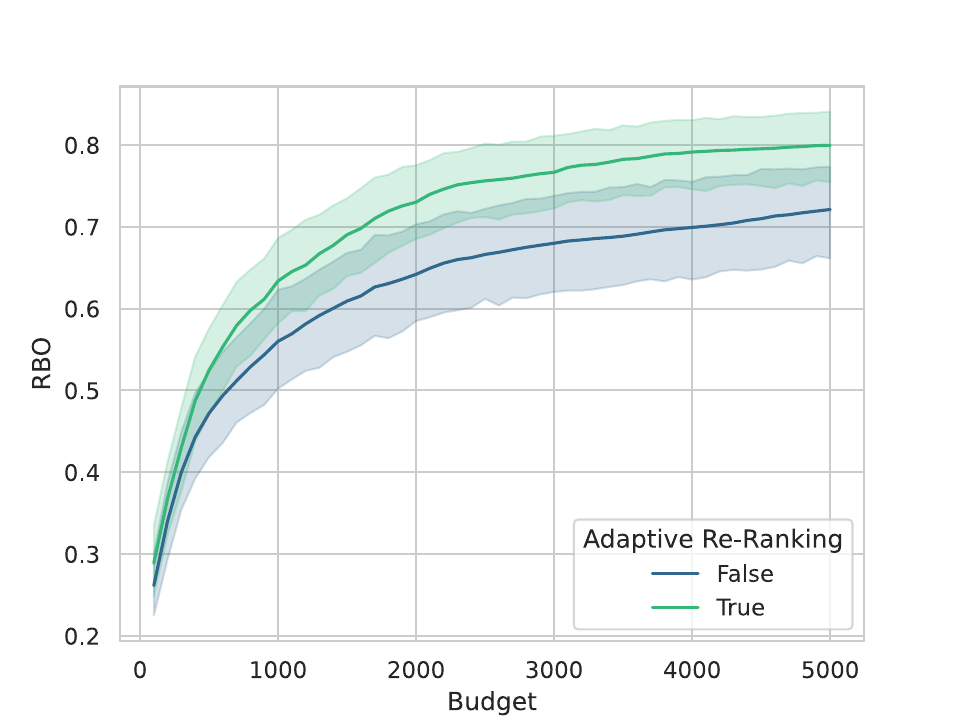}
    }
   \subfloat[][nDCG@10 Performance]{\includegraphics[width=0.5\textwidth]{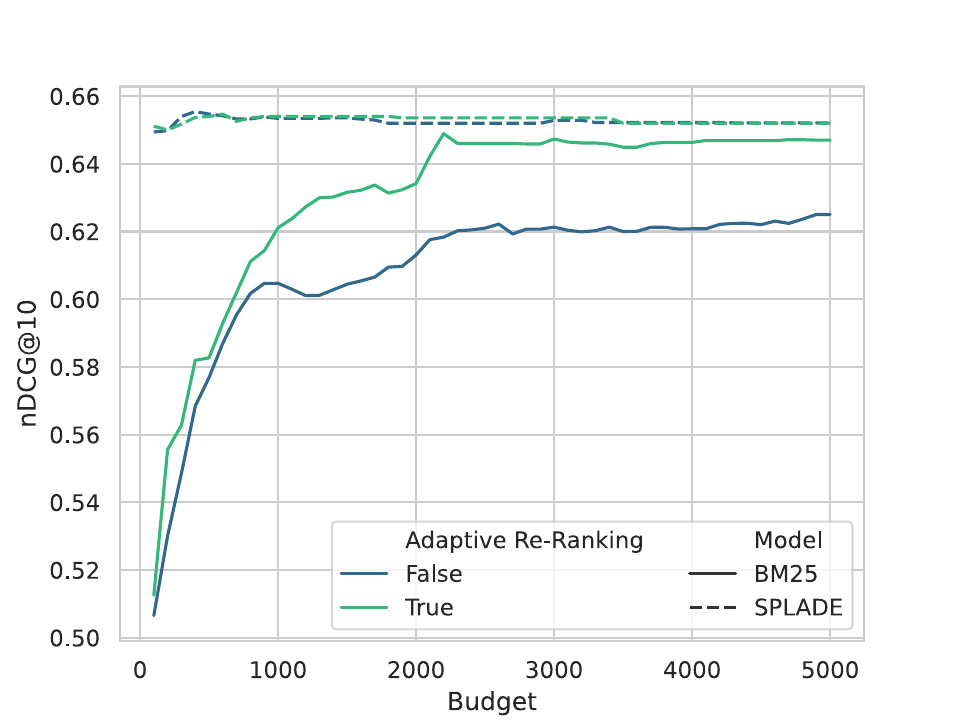}
    }
    \caption{Contrasting adaptive re-ranking over BM25 and SPLADE first stages with full re-ranking at different budgets on TREC Deep Learning 2022 queries. Rankings are truncated to the top 100 results. Post-ranking, duplicates were added to the rankings to follow the current TREC procedure on MS MARCO-v2. Error bands are omitted from Figure (b) for clarity.}
    \label{fig:gar}
\end{figure*}

Observing similar performance (identical top 10 texts for multiple test queries) when adaptive re-ranking is applied with a large corpus graph and budget regardless of first-stage ranker (\textsf{uogtr\_be\_gb} and \textsf{uogtr\_se\_gb}), we consider that by using GAR, the need for complex first-stage retrieval may be reduced. As relevance judgements are not available at the time of analysis, we instead use Deep Learning 2022 relevance judgements and test queries for this analysis. To assess the effect of compute budget on GAR, we apply adaptive ranking with a compute budget ranging from 100 to 5000 texts. Following TREC Deep Learning assessments, we truncate the final rankings to 100 texts. We contrast GAR with full re-ranking, allowing the first-stage retriever to rank up to 5000 texts.

We observe in Figure \ref{fig:gar} (a) that rankings by Rank Biased Overlap (RBO) become increasingly correlated as budget increases, with the trend plateauing around a budget of 4000. Around a budget of 1000, variance begins to increase as rankings converge, whereas, below this point, BM25 and SPLADE rankings are almost completely diverging. As one would expect, the correlation between full re-rankings of BM25 and SPLADE is not as strong, however does follow a similar trend to adaptive re-ranking.

We also observe that from a budget of 2000, it can be seen in Figure \ref{fig:gar} (b) that BM25 and SPLADE first stages are within a small margin of each other in terms of nDCG@10, we found that on both P@10 and R@100, a similar trend was present. We observe that SPLADE is only marginally improved by adaptive re-ranking, as is shown by contrast with full re-ranking and performance measured on the Deep Learning 2023 test queries (\textsf{uogtr\_se} versus \textsf{uogtr\_se\_gb}) in Table \ref{tab:main}. We observe that full re-ranking of BM25 fails to show continuous improvement with nDCG@10 performance plateauing around a budget of 2000, similar to adaptive re-ranking, however, at a significantly lower value (t-test, p < 0.05). This can be attributed to lexical mismatch as full re-ranking is limited by term overlap with a query, whereas GAR can make document-wise term overlap comparisons. We find that metric performance between SPLADE and BM25 using adaptive re-ranking is insignificant from a budget of 2000 (t-test, p < 0.05). Addressing RQ (3), in cases where labelled data is unavailable or financially infeasible to collect, this finding is compelling as the performance of a learned sparse model can be closely replicated by BM25 when using adaptive re-ranking with the caveat that one must have a suitable re-ranker.

\section{Conclusions}
\label{sec:conclus}

In summary, our participation in the TREC 2023 Deep Leaning track has been insightful in validating recent approaches to the expansion of both query terms and first-stage rankings. In answering our research questions, we have found that generative relevance feedback can transfer to a monoELECTRA cross-encoder and is further bolstered by adaptive re-ranking. We find that though generative relevance feedback can be generally effective, the approach is sensitive to the form of the query being direct as opposed to conversational, in which case performance can degrade due to artefacts from instruction-tuning. We also find that with sufficient compute budget and corpus graph size, a first-stage lexical model can closely replicate the metric performance of a learned sparse retrieval model, with both models' rankings becoming increasingly correlated by RBO reaching a maximum correlation of 0.80.

\bibliographystyle{ACM-Reference-Format}
\bibliography{sample-base}

\end{document}